\documentclass[a4paper,12pt]{article}
\usepackage{epsfig}
\usepackage{rotating}

\def\gsim{\raise0.3ex\hbox{$\;>$\kern-0.75em\raise-1.1ex\hbox{$\sim\;$}}}

\def\nn{\nonumber}

\newcommand{\lsim}{\raise0.3ex\hbox{$\;<$\kern-0.75em\raise-1.1ex\hbox{$\sim\;$}}}

\newcommand{\chipm}[1]{\tilde \chi^\pm_{#1}}
\newcommand{\snu}[1]{\tilde \nu_{#1}}

\newcommand{\mchip}[1]{m_{\tilde \chi^+_{#1}}}

\newcommand{\msnu}[1]{m_{\tilde \nu_{#1}}}

\newcommand{\AddrAHEP}{%
 \it AHEP Group, Instituto de F\'{\i}sica Corpuscular --
  C.S.I.C.\\ Universitat de Val{\`e}ncia \\
  Edificio de Institutos de Paterna, Apartado 22085,
  E--46071 Val{\`e}ncia, Spain\\}

\newcommand{\AddrETH}{%
 \it Institut f\"ur Theoretische Physik, Universit\"at Z\"urich, \\ 
CH-8057 Z\"urich, Switzerland}

\begin{document}

\begin{flushright}hep-ph/0409241\\  IFIC/04-50\\  ZU-TH 15/04 \\ \today
  \end{flushright}  

\begin{center}  
  \textbf{\large 
R-parity violating sneutrino decays}\\[10mm]

{D. Aristizabal Sierra${}^1$, M. Hirsch${}^1$ and W. Porod${}^{1,2}$ } 
\vspace{0.3cm}\\ 

$^1$ \AddrAHEP
$^2$  \AddrETH
\end{center}


\begin{abstract}
R-parity can be violated through either bilinear and/or trilinear terms in 
the superpotential. The decay properties of sneutrinos can be 
used to obtain information about the relative importance of these couplings 
provided sneutrinos are the lightest supersymmetric particles. We show 
that in some specific scenarios it is even possible to decide whether 
bilinear or trilinear terms give the dominant contribution to the neutrino 
mass matrix.
\end{abstract}

\section{Introduction}

Supersymmetry offers many possibilities to describe the observed neutrino 
data. The most popular one is certainly the usual seesaw mechanism, which 
introduces heavy right-handed neutrinos carrying a $\Delta L=2$ lepton 
number violating Majorana mass. However, there also exists one interesting 
option which is intrinsically supersymmetric, namely, breaking of R-parity. 

The breaking of R-parity can be realized by introducing explicit R-parity 
breaking terms \cite{Hall:1983id} or by a spontaneous break-down of lepton 
number \cite{Romao:1992vu}. The first class of models can be obtained
in mSugra scenarios where depending on the choice of discrete symmetries
various combinations of R-parity violating parameters
are present at the GUT or Planck scale \cite{Allanach:2003eb}. 
The latter class of models leads after electroweak symmetry 
breaking to effective terms, the so-called bilinear terms, which are a 
sub-class of the terms present in the models with explicit R-parity breaking. 
These bilinear terms have an interesting feature: They do not introduce 
trilinear terms when  evolved from one scale to another with 
renormalization group equations (RGEs). 
In contrast, trilinear terms do generate bilinear
terms when evolved from one scale to another.\footnote{An exception is the 
case when also $\mu$ and $B$ identically vanish at this scale. However, 
this is phenomenologically unacceptable.} 
Thus, from the model-building point 
of view it is an interesting question whether there are observables which 
are sensitive to the presence/absence of the different terms.
In this letter we  tackle this question in scenarios where the
sneutrinos are the lightest supersymmetric particles (LSPs) 
taking the
Minimal Supersymmetric Standard Model (MSSM) augmented with bilinear
and trilinear lepton-number and hence R-parity violating couplings as
a reference model. 

One might ask, if there are high-scale models which lead to such scenarios. 
In mSugra motivated scenarios one usually finds that
either the lightest neutralino or the  right stau is
the LSP. 
In the case of mSugra with R-parity violation it has been shown
in ref.~\cite{Allanach:2003eb} that sneutrinos can be the LSP provided
the R-parity violating parameters follow some specific structures and 
are sufficiently large. Also in models with conserved R-parity there
are some regions in the mSugra parameter space where the sneutrinos can 
be LSPs. However, these scenarios imply $m_0,m_{1/2} < m_W$ and 
small $\tan\beta$ and are, thus, excluded by LEP and TEVATRON data 
\cite{Eidelman:2004wy}. 
It is probably this theoretical prejudice why the possibility
of sneutrinos being LSPs has been largely ignored. However, many
models which depart from strict mSugra in one way or another can be
found in the literature. Just to mention a few representative
examples, there are string inspired models where supersymmetry
breaking is triggered not only by the dilaton fields but also by
moduli fields
\cite{Brignole:1993dj}. In $SO(10)$ or $E(6)$ models, where all neutral
gauge bosons, except those forming $Z$ and $\gamma$, have masses 
of the order of $m_{\rm GUT}$ one expects additional D-term contributions
to the sfermion mass parameters at $m_{\rm GUT}$ \cite{Kolda:1995iw}. 
This is equivalent to assuming non-universal values of $m_0$ for 
left sleptons, right sleptons, left squarks and right squarks. Thus 
sneutrinos can easily be the LSPs in these models.
In the following we will take the general MSSM at the electroweak scale 
as a model as we do not want to rely on specific features of a particular
high scale model.

It has been shown that in models where R-parity violation is the source 
of neutrino masses and mixings the decay properties of the LSP are 
correlated to neutrino properties 
\cite{Mukhopadhyaya:1998xj,Porod:2000hv,Restrepo:2001me,Hirsch:2002ys%
,Chun:2002rh,Hirsch:2003fe,Bartl:2003uq,Jung:2004rd}. In this letter
we will work out some general features of these correlations for
sneutrinos and we will discuss how they depend on the dominance of
various couplings, either bilinear or trilinear R-parity violating
couplings. 
The scenarios we consider are:
(i) a scenario where the contributions to the neutrino mass matrix are
dominated by bilinear terms, (ii) a scenario where bilinear terms and
trilinear terms are of equal importance and (iii) a scenario where
trilinear couplings give the dominant contributions to neutrino masses.

Limits on sneutrinos decaying through R-parity violating operators have 
been published by all LEP collaborations
\cite{Heister:2002jc,Achard:2001ek,Abbiendi:2003rn,delphi} with
similar results.  Limits found for sneutrino LSPs range from
approximately $m_{\tilde \nu} \ge 82-100$ GeV for the different
sneutrino flavours and different analyses.
In ref.~\cite{Abdallah:2002wt} larger mass limits up to $200$ GeV have
been reported considering single sneutrino production in an s-channel
resonance in $e^+e^-$ collisions
\cite{Dimopoulos:1988jw,Barger:1989rk}. However, these limits are
irrelevant for us because they apply only for large sneutrino widths.
In our case we find sneutrino widths of the order of eV.

This letter is organized as follows: in Sect.~\ref{sec:model} we
discuss briefly the model considered, the choice of basis as well as
aspects related to neutrino physics.  In Sect.~\ref{sec:prod} we
discuss sneutrino production at future colliders, such as the LHC and a
future $e^+ e^-$ collider. In addition we discuss general aspects
concerning sneutrino signatures. In Sect.~\ref{sec:num} we discuss
specific scenarios pointing out how to obtain information on the relative
importance of the various couplings.
Finally we present in
Sect.~\ref{sec:conc} our conclusions.

\section{Model}
\label{sec:model}

\subsection{Generalities}

The most general form for the R-parity violating and lepton number 
violating part of the superpotential is given by
\begin{equation}
W_{R_p \hspace{-0.8em}/\;\:}=\varepsilon_{ab}\left[ 
\frac{1}{2}\lambda_{ijk}\widehat L_i^a\widehat L_j^b\widehat E_k 
+\lambda'_{ijk}\widehat L_i^a\widehat Q_j^b\widehat D_k 
+\epsilon_i\widehat L_i^a\widehat H_u^b\right]~.
\label{eq:rpvpot}
\end{equation}
In addition, for consistency, one has to add three more bilinear terms 
to $V_{soft}$, the SUSY breaking potential, 
\begin{equation}
V_{\rm{soft},R_p \hspace{-0.8em}/\;\:} = -\varepsilon_{ab}
          B_i\epsilon_i\widetilde L_i^aH_u^b 
+ \frac{1}{2} A_{\lambda,ijk} \lambda_{ijk}\tilde L_i^a\tilde L_j^b\tilde E_k 
+A_{\lambda',ijk}\lambda'_{ijk}\tilde L_i^a\tilde Q_j^b\tilde D_k 
\, .
\label{eq:rpvsoft}
\end{equation}
Eq.~(\ref{eq:rpvpot}) contains in general 36 different $\lambda_{ijk}$, 
and $\lambda'_{ijk}$ and three $\epsilon_i$, for a total of 39 parameters. 
As already pointed out in \cite{Hall:1983id}, the number 
of parameters in Eq.~(\ref{eq:rpvpot}) can be reduced by 3 by a suitable 
rotation of basis and a number of authors have used this freedom 
to absorb the bilinear parameters $\epsilon_i$ into re-defined 
trilinear parameters \cite{dreiner:1997uz} (for connections to mSUGRA models
 see also \cite{Allanach:2003eb}). It is important to note 
that in general there is no rotation which can eliminate 
the bilinear terms in Eq.~(\ref{eq:rpvsoft}) and Eq.~(\ref{eq:rpvpot}) 
simultaneously \cite{Hall:1983id}. The only exception is the case
where $B = B_i$ and $m_{H_d}^2=m_{L_i}^2$ for all $i$. 
However, these conditions are not stable under RGE running 
\cite{decarlos:1997yh}. 
Therefore we are back to 39 independent parameters relevant for
neutrino physics in the most general case. The 36 A-parameter contribute
only at the 2-loop level to neutrino physics and are ignored in the following
as they also do not affect the sneutrino decays in our scenarios.

The first term in Eq.~(\ref{eq:rpvsoft}) induces non-zero vacuum expectation 
values  $v_{L_i}$ for the sneutrinos. These are connected to the parameters 
$B_i$ via the tadpole equations \cite{hirsch:2000ef,h2} and thus are 
not independent. We will trade the $B_i$ for the $v_{L_i}$ as 
independent parameters in the following.

\subsection{Basis rotation}

Although in the numerical part of this work we always diagonalize 
all mass matrices exactly starting from the general basis, 
Eq.~(\ref{eq:rpvpot}), for an analytical understanding of our results 
it is very useful to define an ``$\epsilon$-less'' basis using 
the following transformation.
The bilinear terms can be removed from the superpotential by a 
series of three rotations\footnote{For alternative forms of this
rotation see ref.~\cite{Allanach:2003eb,hirsch:2000ef}, for the
case of complex parameters see ref.~\cite{Dreiner:2003hw}.}
 ${\bf R} = {\bf R_3} {\bf R_2} {\bf R_1}$:
\begin{eqnarray}
{\bf R} = 
\left(\begin{array}{ccccc}
c_1 c_2 c_3 & -c_2 c_3 s_1 & -c_3 s_2 & -s_3 \cr
s_1         &  c_1         &        0 & 0 \cr
c_1 s_2     & -s_1 s_2     & c_2      & 0 \cr
c_1 c_2 s_3 & -c_2 s_1 s_3 & -s_2 s_3 & c_3  
\end{array}\right) \, \, , 
\label{defrotR}
\end{eqnarray}
where the angles $\theta_i$ are defined by
\begin{equation}
\sin\theta_3=\frac{\epsilon_3}{\mu'} \hskip5mm
\sin\theta_2=\frac{\epsilon_2}{\mu''} \hskip5mm
\sin\theta_1=\frac{\epsilon_1}{\mu'''} \hskip5mm ,
\label{defang}
\end{equation}
and $\mu'= \sqrt{\mu^2+\epsilon_3^2}$, $\mu''= \sqrt{\mu^{'2}+\epsilon_2^2}$ 
and $\mu'''= \sqrt{\mu^{''2}+\epsilon_1^2}$. Note that with 
$\epsilon_i \ll \mu$, as required by neutrino data, 
$\mu'\simeq \mu'' \simeq \mu''' \simeq \mu$ to a very good 
approximation. After this rotation the trilinear parameters in 
Eq.~(\ref{eq:rpvpot}) get additional contributions. Assuming, without 
loss of generality, that the lepton and down type Yukawa couplings are 
diagonal\footnote{After performing 
the rotation in Eq.~(\ref{defrotR}) one has to perform a rotation 
of the right-handed leptons to keep the lepton Yukawa matrix diagonal.} 
they are given to leading order in $\epsilon_i/\mu$ as :
\begin{equation}
\lambda'_{ijk} \to \lambda'_{ijk} + \delta_{jk} h_{d_k} \frac{\epsilon_i}{\mu}
\label{efflamp}
\end{equation}
and
\begin{eqnarray}
\lambda_{ijk} \to \lambda_{ijk} +  \delta \lambda_{ijk}  , \hskip42mm
\label{efflam} \\
\delta \lambda_{121} = h_{e} \frac{\epsilon_2}{\mu}, \hskip5mm
\delta \lambda_{122} = h_{\mu} \frac{\epsilon_1}{\mu}, \hskip5mm
\delta \lambda_{123} = 0 \nonumber \\ \nonumber
\delta \lambda_{131} = h_{e} \frac{\epsilon_3}{\mu}, \hskip5mm
\delta \lambda_{132} = 0, \hskip5mm
\delta \lambda_{133} = h_{\tau} \frac{\epsilon_1}{\mu} \\ \nonumber 
\delta \lambda_{231} = 0, \hskip5mm
\delta \lambda_{232} = h_{\mu} \frac{\epsilon_3}{\mu}, \hskip5mm
\delta \lambda_{233} = h_{\tau} \frac{\epsilon_2}{\mu} \\ \nonumber 
\end{eqnarray}
The essential point to notice is that the additional contributions 
in Eqs.~(\ref{efflamp}) and (\ref{efflam}) follow the hierarchy 
dictated by the down quark and charged lepton masses of the 
standard model. 

\subsection{Neutrino masses}

Contributions to the neutrino mass matrix are induced at tree-level by
the bilinear R-parity violating terms. In the ``$\epsilon$-less'' basis 
they can be expressed as \cite{Joshipura:ib,hirsch:1998kc}
\begin{eqnarray} \label{eq:meff}
m_{\rm eff}=\frac{(M_1 g^2+M_2 {g'}^2)\mu^2}
{4\ \det({\mathcal M}_{\chi^0})}
\left(\hskip -2mm \begin{array}{ccc}
v_e^2 & v_e v_{\mu} & v_e v_{\tau} \\ 
v_e v_{\mu} & v_{\mu}^2 & v_{\mu} v_{\tau} \\ 
v_e v_{\tau} & v_{\mu}v_{\tau} & v_{\tau}^2 
\end{array} \right),
\end{eqnarray}
Only one neutrino acquires mass at the tree level from 
Eq. (\ref{eq:meff}). Therefore $m_{\rm eff}$ is diagonalized 
with only two mixing angles which can be expressed in terms of 
$v_i$:
\begin{eqnarray} \label{eq:choozangle}
\tan\theta_{13} &=& - \frac{v_e}
                   {\sqrt{(v_{\mu}^2+v_{\tau}^2)}}~, \\
 \label{eq:atmangle}
\tan\theta_{23} &=& -\frac{v_{\mu}}{v_{\tau}}~.
\end{eqnarray}
One-loop contributions to the neutrino mass matrix from bilinear terms
have been discussed extensively, see for example \cite{hirsch:2000ef,h2} 
and will not be repeated here. 
The contributions from trilinear terms are given by 
$m^{1-loop}=m^{\lambda}+ m^{\lambda'}$, where 
\cite{Allanach:2003eb,Bartl:2003uq,Borzumati:2002bf}
\begin{equation} \label{eq:Triloopmass}
m_{ii'}^{\lambda (\lambda')}=-\frac{1(N_c)}
{32\pi^2}~\lambda^{(')}_{ijk}\lambda^{(')}_{i'kj}
\left[
m_k\sin2\theta_j \ln\left(\frac{m^2_{2j}}{m^2_{1j}}\right)+
m_j\sin2\theta_k \ln\left(\frac{m^2_{2k}}{m^2_{1k}}\right)
\right]~,
\end{equation}
where $m_k$ is the appropriate fermion mass, $\theta_j$ denotes the
appropriate sfermion mixing angle and $m^2_{(1,2)j}$ are the
corresponding sfermion masses. Note the manifest symmetry of this
matrix as well as the presence of logarithmic factors.
We stress that Eq. (\ref{eq:Triloopmass}) is an approximation to the
full 1-loop calculation. The complete formula including bilinear and trilinear
couplings can be easily obtained by replacing the following  couplings
in Ref.~\cite{hirsch:2000ef}:
\begin{eqnarray}
O^{cns}_{Lijk} &\to& O^{cns}_{Lijk}
         - \lambda_{rst} V^*_{i,2+t} N^*_{j,4+r} R^{S^\pm}_{k,2+s} \\
O^{cns}_{Rijk} &\to& O^{cns}_{Rijk}
         - \lambda_{rst} U_{i,2+t} N_{j,4+r} R^{S^\pm}_{k,5+s}  
\end{eqnarray} 
in Eqs.~(B4)--(B6) and
\begin{eqnarray}
 O^{dns}_{Lijk} &\to&  O^{dns}_{Lijk}
            - \lambda'_{rst} N^*_{j,4+r} R^{\tilde d *}_{k,s} R^d_{Ri,t} \\
 O^{dns}_{Rijk} &\to&  O^{dns}_{Rijk}
            - \lambda'_{rst} N_{j,4+r} R^{\tilde d *}_{k,t+3} R^{d*}_{Li,s} 
\end{eqnarray} 
in Eqs.~(B12)--(B14).
In the numerical results below we use the complete formula obtained by these
replacements.

\section{Sneutrino pair production and decays}
\label{sec:prod}

\begin{figure}
\setlength{\unitlength}{1mm}
\begin{picture}(160,70)
\put(-3,0){\mbox{\epsfig{figure=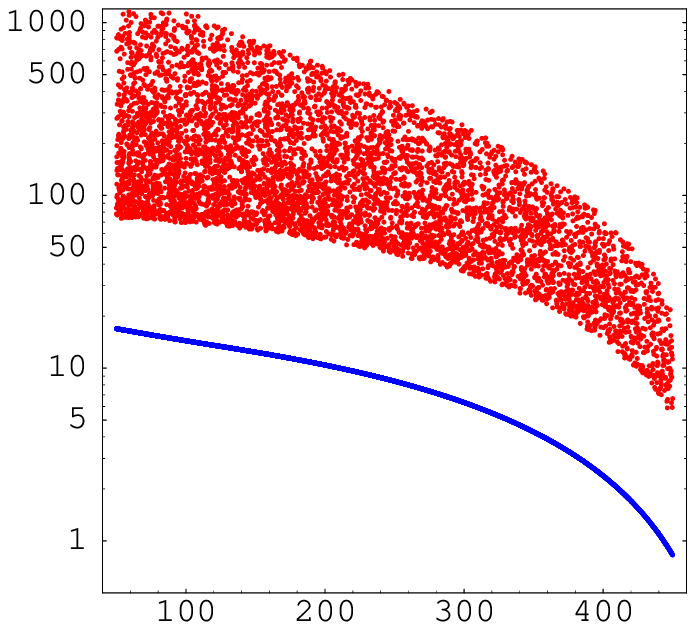,height=6.8cm,width=6.8cm}}}
\put(-5,67){\makebox(0,0)[bl]{ a)}}
\put(5,66){\makebox(0,0)[bl]{
       $\sigma(e^+ e^- \to \tilde \nu_i \tilde \nu_i)$~[fb]}}
\put(66,-1){\makebox(0,0)[br]{$m_{\tilde \nu}$ [GeV]}}
\put(50,50){\makebox(0,0)[br]{\small $\snu{e} \snu{e}$}}
\put(29,24){\makebox(0,0)[br]{
   \small $\snu{\mu} \snu{\mu}$, $\snu{\tau} \snu{\tau}$ }}
\put(70,0){\mbox{\epsfig{figure=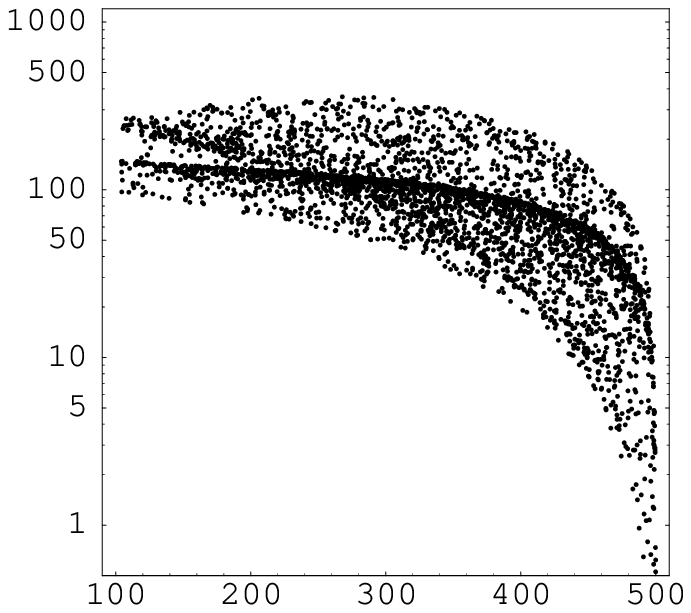,height=6.8cm,width=6.8cm}}}
\put(68,67){\makebox(0,0)[bl]{ b)}}
\put(78,66){\makebox(0,0)[bl]{
    $\sigma(e^+ e^- \to \tilde \chi^+_1 \tilde \chi^-_1)$~[fb]}}
\put(138,-2){\makebox(0,0)[br]{$m_{{\tilde \chi}^-_1}$~[GeV]}}
\end{picture}
\caption[]{ Production cross section for a) sneutrinos and b) charginos
at an 1 TeV $e^+ e^-$ collider with unpolarized beams.}
\label{fig:prod}
\end{figure}

In the following we discuss sneutrino production at a future $e^+ e^-$
collider and the LHC as well as sneutrino decays. In
Fig.~(\ref{fig:prod}) we show the cross sections at an 1 TeV $e^+ e^-$
collider with unpolarized beams. Here we have scanned the following
parameter range: $95 \le m_{\tilde \nu} \le 500$~GeV, $-1000 \le \mu
\le 1000$~GeV, $100 \le M_2 \le 750$~GeV, $3 \le \tan\beta \le 40$. 
We have fixed $M_1$ by the GUT relation $M_1 = 5 \tan^2 \theta_W M_2/3$
to reduce the number of free parameters. This choice does not affect any 
of our conclusions, as only the overall normalization in the tree-level 
neutrino mass matrix in Eq.~(\ref{eq:meff}) is slightly affected. 
In Fig.~(\ref{fig:prod}a) the cross sections for sneutrino production are
shown as a function of $m_{\tilde \nu}$, the points for the lower line
are the cross sections for $\mu$ and $\tau$ sneutrinos whereas the
ones in the upper region are for $e$ sneutrinos.  In
Fig.~(\ref{fig:prod}b) the cross section for chargino production is
shown as a function of $m_{{\tilde \chi}^-_1}$. Here we have taken
into account only those points where the chargino is heavier than the
sneutrinos and is heavier than 104~GeV to be consistent with the LEP
constraint \cite{Eidelman:2004wy}.
For the direct production we see in Fig.~(\ref{fig:prod}a)
that up to a mass of $\sim$450 GeV the
electron sneutrino has a cross section larger than 10 fb. Assuming
an integrated luminosity of 1~ab$^{-1}$ this implies that at least
$10^4$ sneutrino pairs are produced directly. 

The sneutrino has the following decay modes
\begin{eqnarray}
 \tilde \nu_i &\to& q_j \, \bar{q}_k  \\
 \tilde \nu_i &\to& l^+_j \, l^-_k \\
 \tilde \nu_i &\to& \nu_j \, \nu_k 
\end{eqnarray}
where the hadronic final states contain in general $d$-type quarks.
As we will see below,  measurable relations to neutrino mixing
angles can be obtained mainly by considering final states with charged leptons.
Since one expects sneutrinos of different flavours to be nearly mass
degenerate it will probably not be possible to decide which
sneutrino flavour has been produced in the direct production.
As has been shown in \cite{Hirsch:2003fe} the sharpest correlations between
sneutrino decay properties and neutrino mixing angles are obtained if
the sneutrino flavour is known. Therefore we propose to study sneutrinos
stemming from chargino decays
\begin{eqnarray}
\chipm{1} \to \snu{l} l^\pm
\label{eq:charginodecay}
\end{eqnarray}
and use the additional lepton as a flavour tag. Clearly one has to
be careful not to confuse this lepton with the leptons stemming from
the sneutrino decays. To minimize the combinatorial background one has
to know the sneutrino and chargino masses. At an LC this information
can be obtained via threshold scans. At the LHC one can in principal
proceed in the following way to obtain the necessary information:
(i) Take the data sample containing only two leptons as they will
contain events stemming from the pair production of $\tilde q_R$ where
$\tilde q_R$ decays according to
\begin{eqnarray}
 \tilde q_R \to q \tilde \chi^0_1 \to q \nu \tilde \nu  \,\, .
\label{eq:chainR}
\end{eqnarray}  
This sample now contains  events where only one of the sneutrinos decays
into charged leptons and the second one into either neutrinos or jets.
(ii) The chargino mass can be obtained 
by a suitable adjustment of the so-called 'edge' variables
discussed in ref.~\cite{Allanach:2000kt} which are obtained by studying
invariant momentum spectra of leptons and quarks. Interesting decay
chains are for example those containing
\begin{eqnarray}
\tilde q_L \to \chipm{1} q' \to \snu{l} l^\pm q'    \,\, .
\label{eq:chainL}
\end{eqnarray}
Here one should first consider events containing three charged leptons,
which occurs for example if one of the ${\tilde q}_L$ decays according to the
chain in Eq.~(\ref{eq:chainL}) and the second one according to the chain
in Eq.~(\ref{eq:chainR}). Clearly one has in this case a combinatorial
background which has two sources: (i) the sneutrino from the chargino
decay decays hadronically or invisible and the second one into charged
leptons and (ii) the pairing of the charged leptons with the same charge
is incorrect. In the latter case a study of the kinematics shows 
that there is exactly one kinematical situation in the rest frame of
the chargino where this is possible, namely if the angle $\theta$ between
the leptons with the same charge fulfills the following relation:
\begin{eqnarray}
\cos\theta &=& \frac{1 - 2 \beta }{\beta}  \,\, , \,\,
\beta = \frac{\mchip{1}^2 - \msnu{l}^2}{\mchip{1}^2 + \msnu{l}^2} \,\, .
\label{eq:cost}
\end{eqnarray}
The condition $|\cos\theta|\le 1$ requires 
$\msnu{l} / \mchip{1}\lsim 0.755$. For larger ratios no confusion is possible
because the lepton stemming from the chargino has much less energy than
the one from the sneutrino decay. 

In case of four and more leptons in the event, which have to be
analyzed for example
in scenarios where the sneutrinos decay mainly into charged leptons,
more kinematical configurations exists where a confusion of the various
leptons is possible. A principle way to minimize the combinatorial background
is to take out the events with these special kinematical configurations. 
It is clear that this type of analysis is an experimental challenge.
Detailed Monte Carlo studies will be necessary to determine
the impact of the experimental environments of future LHC and $e^+ e^-$ 
collider experiments on these considerations.

We want to note, that in scenarios, where sneutrinos are the
LSPs, the left charged sleptons are not much heavier as the mass difference
 is
roughly given by: $m^2_{\tilde l,L} - m^2_{\tilde \nu} \simeq - \cos 2
\beta m^2_W > 0$. The left sleptons
decay in these scenarios either via three body decays, which conserve
R-parity, into
\cite{Ambrosanio:1997bq,Djouadi:2000bx}
\begin{eqnarray}
 \tilde l_L &\to& \tilde \nu \, \bar{q} q' \,\,\, , \,\,\,
 \tilde l_L \to \tilde \nu \, \nu l       
\end{eqnarray}
or via R-parity violating couplings into
\begin{eqnarray}
 \tilde l_L &\to&  \bar{q} q'  \,\,\, , \,\,\,
 \tilde l_L \to  \nu l \, .       
\end{eqnarray}
The latter decay modes give in principal rise to additional
observables correlated with neutrino physics. However, we have found
that for mass differences larger than $\simeq 5$--10 GeV the three
body decays clearly dominate. Therefore this additional information is
only available if either $\tan\beta$ is small and/or if all particles
have masses above $\gsim 400$~GeV.

At the LHC \cite{Chaichian:2003kf} as well as at a $e^+ e^-$ machine  
\cite{Dimopoulos:1988jw,Barger:1989rk} it is possible 
to produce sneutrinos in s-channel reactions if R-parity is violated. 
Assuming that all R-parity violating couplings are at most as large 
as the largest one compatible with neutrino physics it turns out 
that these processes are strongly suppressed. We therefore neglect them 
in the following considerations. However, if there is an 'anti-hierarchical' 
structure present in the trilinear couplings allowing  e.g.~for 
large $\lambda_{121}$ and/or $\lambda'_{k11}$ couplings this would lead to 
additional statistics as well as to additional interesting information.

\section{Numerical Studies}
\label{sec:num}

We now turn to the discussion of the numerical results. In all 
the following studies we first created a set of supersymmetric 
spectra in which sneutrinos are the LSPs. Instead of resorting 
to some specific supersymmetry breaking scheme, this is simply 
achieved by taking the sneutrino mass as a free parameter, 
as discussed above. After selecting sneutrino masses to obey the LEP 
bounds \cite{Heister:2002jc,Achard:2001ek,Abbiendi:2003rn,delphi} 
 we  add R-parity violating 
parameters taking into account constraints from neutrino physics. 
These are $\Delta m^2_{Atm}$ and $\tan^2\theta_{Atm}$, the 
atmospheric mass squared difference and mixing angle, 
$\Delta m^2_{\odot}$ and $\tan^2\theta_{\odot}$, the 
solar mass squared difference and mixing angle, as well as 
the upper limit on $\sin^22\theta_{R}$, the reactor angle. 
Different assumptions as to which of the R-parity breaking 
parameters give the dominant contributions to the neutrino 
mass matrix give different decay patterns of the sneutrinos, 
which we will now discuss in turn. 
 The scenarios we consider are:
(i) a scenario where the contributions to the neutrino mass matrix are
dominated by bilinear terms, (ii) a scenario where bilinear terms and
trilinear terms are of equal importance and (iii) a scenario where
trilinear couplings give the dominant contributions to neutrino masses. 
One should keep in mind here that neutrino physics does not fix
all of the R-parity violating parameters and, thus, there will 
be several sneutrino decay channels which are not directly related to 
neutrino physics. The main assumption in the discussion below is that
those decays related to neutrino physics have a 
branching ratio in the order of at least per-mille.
Moreover, we will assume that the flavour
of sneutrinos has been tagged with the help of the additional lepton
in the chargino decay. In addition, we note that all R-parity parameters 
discussed below are given in the ``$\epsilon$-less'' basis discussed in 
section~\ref{sec:model}.

\subsection{Scenario I: Bilinear terms only}

This scenario has been discussed previously briefly in \cite{Hirsch:2003fe}. 
Here we assume that the atmospheric mass scale and angle is dominated by the 
tree-level contribution to the mass matrix, whereas the solar mass scale is 
induced by loops. 

In this scenario the trilinear couplings of the sneutrinos to down quarks 
and charged leptons follow a hierarchy dictated by the standard model quark 
and charged lepton masses, see Eqs (\ref{efflamp}) and (\ref{efflam}). With 
this observation and the help of Eqs (\ref{eq:meff}) and 
(\ref{eq:Triloopmass}) the relative size of all sneutrino couplings can be 
fixed with the help of neutrino data. The atmospheric neutrino mass scale 
\footnote{R-parity violating neutrino mass models produce hierarchical 
neutrino spectra.} can be estimated to be $\sqrt{\Delta m^2_{Atm}} \sim 
{\cal O}(0.05)$ eV, while the solar neutrino mass scale is approximately 
given by $\sqrt{\Delta m^2_{\odot}} \sim {\cal O}(0.009)$ eV. From 
Eq. (\ref{eq:meff}) one can estimate $\frac{v}{M_2} \sim 10^{-6}$, 
where $v = \sqrt{\sum_i v_{i}^2}$. On the other hand, $\lambda'_{i33}$ 
can be estimated from Eq. (\ref{eq:Triloopmass}) to be of the order of 
typically $\lambda'_{i33} \sim h_b \frac{\epsilon_i}{\mu} \sim 
{\cal O}(10^{-4})$ to correctly produce the solar neutrino mass 
scale. 

One expects therefore that the most important final state for sneutrinos 
is $b{\bar b}$, independent of the sneutrino generation. Electron and 
muon sneutrinos will decay also to $\tau{\bar \tau}$ final states
with a relative ratio of 
\begin{equation}
\frac{Br({\tilde \nu}_{e,\mu} \rightarrow \tau{\bar \tau})}
     {Br({\tilde \nu}_{e,\mu} \rightarrow b{\bar b})} \simeq \frac{h^2_{\tau}}
                                           {3 h^2_b (1 + \Delta_{QCD})}
\label{bbovertautau}
\end{equation}
independent of all other parameters. Here $\Delta_{QCD}$ are the QCD
radiative corrections. Decays to $\mu{\bar\mu}$ (and non-$b$ jets) final
states are suppressed by the corresponding Yukawa couplings squared.

\begin{figure}[t]
\hskip5mm
\includegraphics[width=65mm,height=5cm]{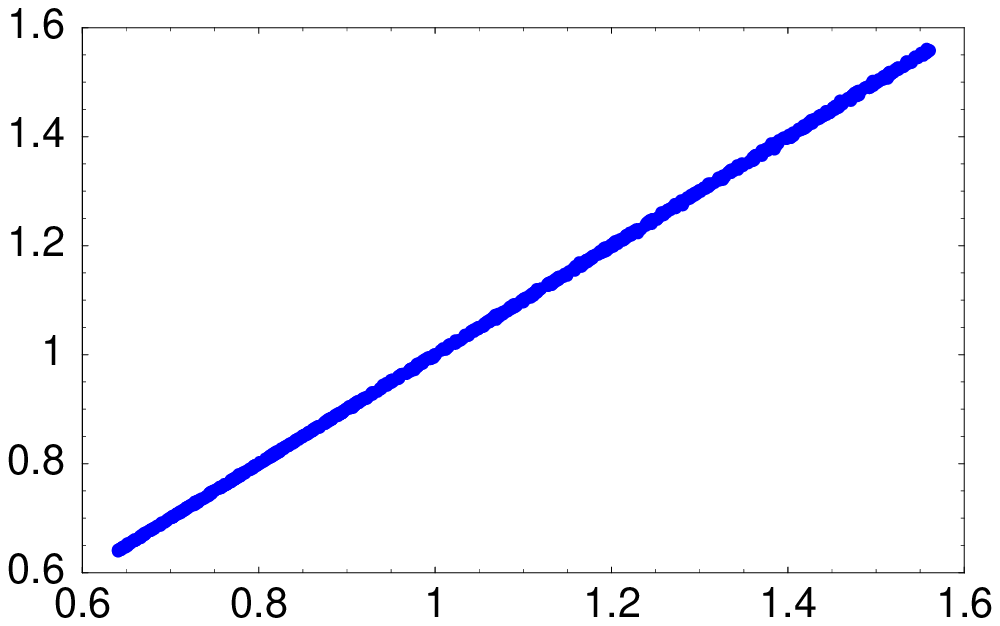}
\hskip5mm
\includegraphics[width=65mm,height=5cm]{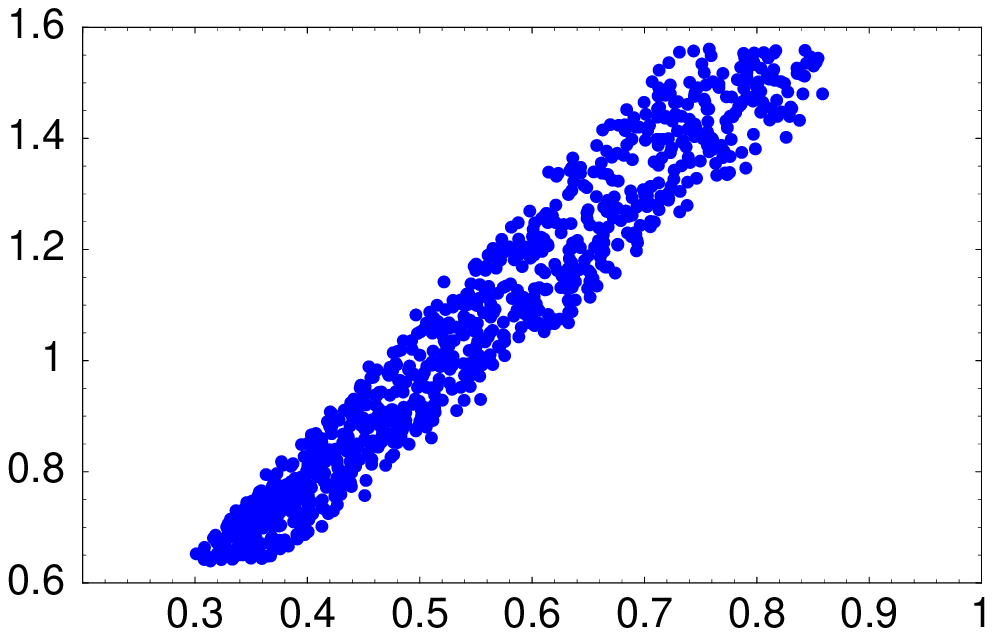}
\vskip-8mm\hskip0mm
\begin{rotate}{90}
$Br({\tilde \nu}_{\tau}\to e\tau)/Br({\tilde \nu}_{\tau}\to\mu\tau)$
\end{rotate}
\vskip-4mm\hskip75mm
\begin{rotate}{90}
$Br({\tilde \nu}_{\tau}\to e\tau)/Br({\tilde \nu}_{\tau}\to\mu\tau)$
\end{rotate}
\vskip2mm
\hskip55mm
$(\epsilon_{1}/\epsilon_{2})^2$ 
\hskip55mm $\tan^2\theta_{\odot}$
\vskip0mm
\caption{Ratio of branching ratios 
$Br({\tilde \nu}_{\tau}\to e\tau)/Br({\tilde \nu}_{\tau}\to\mu\tau)$
versus a) (left) $(\epsilon_{1}/\epsilon_{2})^2$ 
and b) (right) $\tan^2\theta_{\odot}$ for scenario I.}
\label{fig:solbil}
\end{figure}

Tau sneutrinos, on the other hand, will decay to final states 
$e\tau$ and $\mu\tau$ with sizable branching ratios 
\begin{equation}
\frac{Br({\tilde \nu}_{\tau} \rightarrow e\tau)
     [Br({\tilde \nu}_{\tau} \rightarrow \mu\tau)]}
     {Br({\tilde \nu}_{\tau} \rightarrow b{\bar b})} \simeq \frac{h^2_{\tau}}
                                           {3 h^2_b (1+\Delta_{QCD})}
                       \frac{\epsilon_1^2 [\epsilon_2^2]}{\epsilon_3^2}
\label{staucheck}
\end{equation}
The above relation allows one to cross check the consistency of the bilinear 
scenario with neutrino data, as demonstrated in Fig. (\ref{fig:solbil}). 
The current $3\sigma$ allowed range for $\tan^2\theta_{\odot}$ of 
$\tan^2\theta_{\odot} =$ [$0.30,0.59$] fixes 
$Br({\tilde \nu}_{\tau} \rightarrow e\tau)/
Br({\tilde \nu}_{\tau} \rightarrow \mu\tau) \simeq [0.55,1.25]$, 
as can be seen in Fig. (\ref{fig:solbil}).
\footnote{A similar test could be done, in principle, by measuring 
$Br({\tilde \nu}_{\mu} \rightarrow e\mu)/
Br({\tilde \nu}_{\mu} \rightarrow \tau\tau)$, which is also related 
to $\tan^2\theta_{\odot}$. However, the number of events in the 
final state $e\mu$ in ${\tilde \nu}_{\mu}$ decays is suppressed by 
$h_{\mu}^2$.}

Non-zero sneutrino vevs induce the decay ${\tilde \nu} \rightarrow 
\nu\nu$, i.e. by measuring non-zero branching ratios for invisible decays 
one could establish that sneutrino vevs exist. From the estimate on 
$\frac{v}{M_2}$ and $\frac{\epsilon}{\mu}$ discussed above one can estimate 
that branching ratios of sneutrino decays to invisible states should be 
of the order ${\cal O}(10^{-4})$.

\begin{figure}[t]
\begin{center}
\includegraphics[width=75mm,height=5cm]{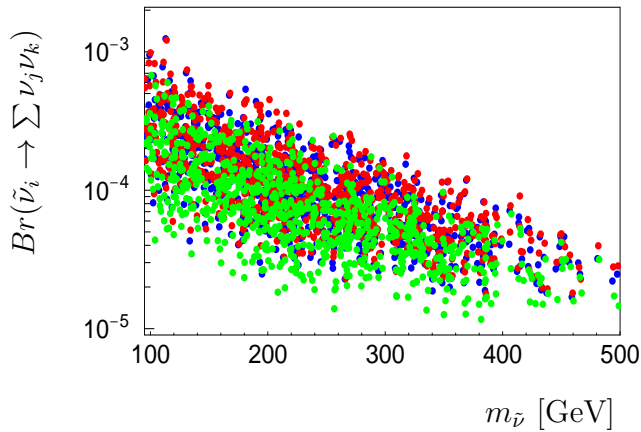}
\end{center}
\vskip-25mm\hskip25mm
\begin{rotate}{90}
$Br(\tilde{\nu}_i \to \sum \nu_j \nu_k)$
\end{rotate}
\vskip17mm
\hskip85mm
$m_{\tilde \nu}$ [GeV]
\vskip0mm
\caption[]{Invisible sneutrino decay branching ratio versus sneutrino mass 
for scenario I. Light (medium, dark) points (green, 
red, blue) are for ${\tilde \nu}_e$ (${\tilde \nu}_{\mu}$, 
${\tilde \nu}_{\tau}$).}
\label{fig:inv}
\end{figure}

Figure (\ref{fig:inv}) shows the calculated branching ratios for 
invisible final states, $Br(\tilde{\nu}_i \to \sum \nu_j \nu_k)$, 
as a function of the sneutrino mass. The figure shows that 
the estimate discussed above is correct within an order of 
magnitude. It also demonstrates that for sneutrinos below 
$m_{\tilde \nu} \le 500$ GeV one expects 
$Br(\tilde{\nu}_i \to \sum \nu_j \nu_k) \ge 10^{-5}$, i.e. with 
the cross sections calculated, see Fig.~\ref{fig:prod}, a few 
events per year are expected with the signature $l^+l^-b{\bar b}$ + 
missing energy (from chargino pair production) and $b{\bar b}$ + missing 
energy (from sneutrino pair production). 

\begin{figure}[t]
\begin{center}
\includegraphics[width=75mm,height=5cm]{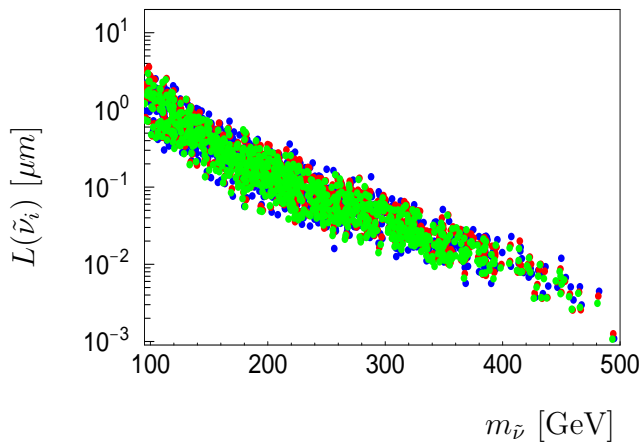}
\end{center}
\vskip-25mm\hskip25mm
\begin{rotate}{90}
$L(\tilde{\nu}_i)$ [$\mu m$]
\end{rotate}
\vskip17mm
\hskip85mm
$m_{\tilde \nu}$ [GeV]
\vskip0mm
\caption[]{Sneutrino decay length versus sneutrino mass 
for scenario I.}
\label{fig:len}
\end{figure}

To measure absolute values of R-parity violating parameters it would
be necessary to measure the decay widths of the 
sneutrinos. Given the current neutrino data, however, such a
measurement seems to be out of reach for the next generation of
colliders. Figure (\ref{fig:len}) shows calculated decay lengths,
assuming a center of mass energy of $\sqrt{s} =1$ TeV, versus 
sneutrino mass. The decay lengths are short compared to sensitivities
expected at a future linear collider which are of order 10 $\mu$m
\cite{Behnke:2001qq}. One can turn this argument around to
conclude that observing decay lengths much
larger than those shown in Fig. (\ref{fig:len}) would rule out
explicit R-parity violation as the dominant source of neutrino mass.

\subsection{Scenario II: $\lambda'_{ijk} \simeq 0$}

\begin{figure}[t]
\hskip10mm
\includegraphics[width=65mm,height=5cm]{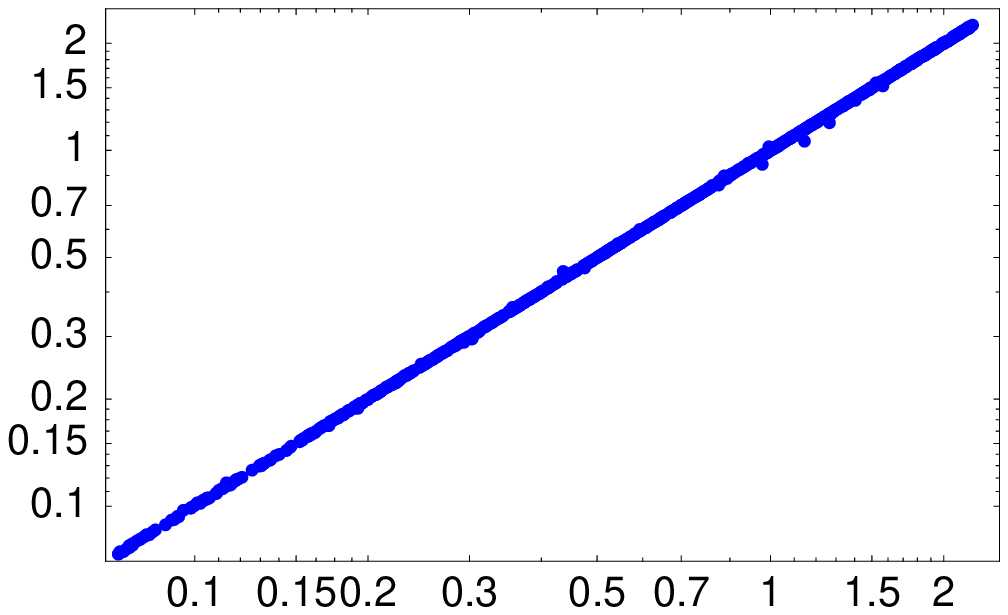}
\includegraphics[width=65mm,height=5cm]{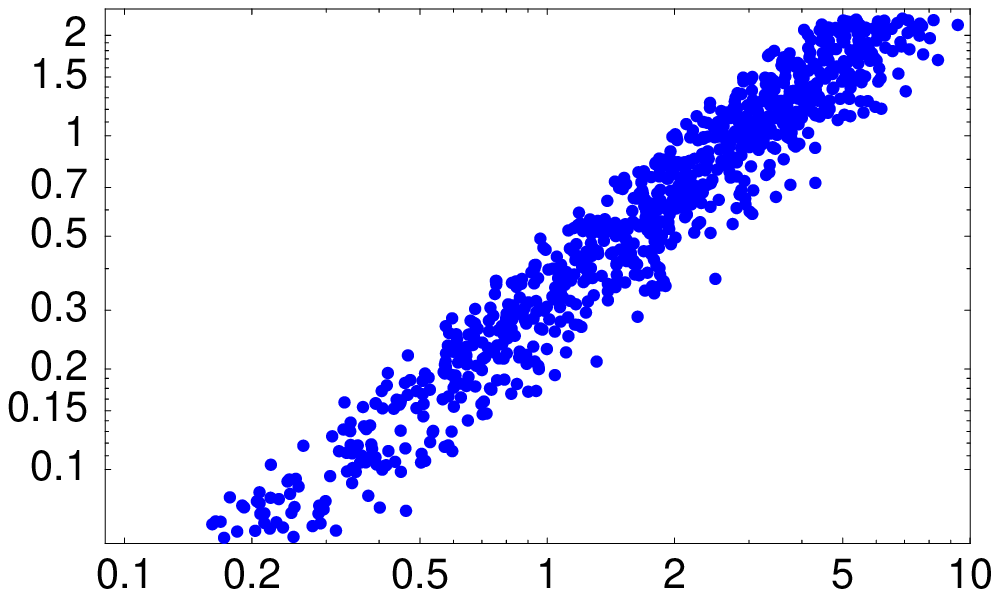}

\vskip-18mm\hskip5mm
\begin{rotate}{90}
$\frac{
Br({\tilde \nu}_e\to\tau\tau)/Br({\tilde \nu}_{\mu}\to\tau\tau)}
{Br({\tilde \nu}_e\to\mu\mu)/Br({\tilde \nu}_{\mu}\to e\mu)}$ 
\end{rotate}
\vskip12mm
\hskip55mm
$(\lambda_{133}/\lambda_{233})^2$ 
\hskip50mm $\tan^2\theta_{\odot}$
\vskip0mm

\hskip10mm
\includegraphics[width=65mm,height=5cm]{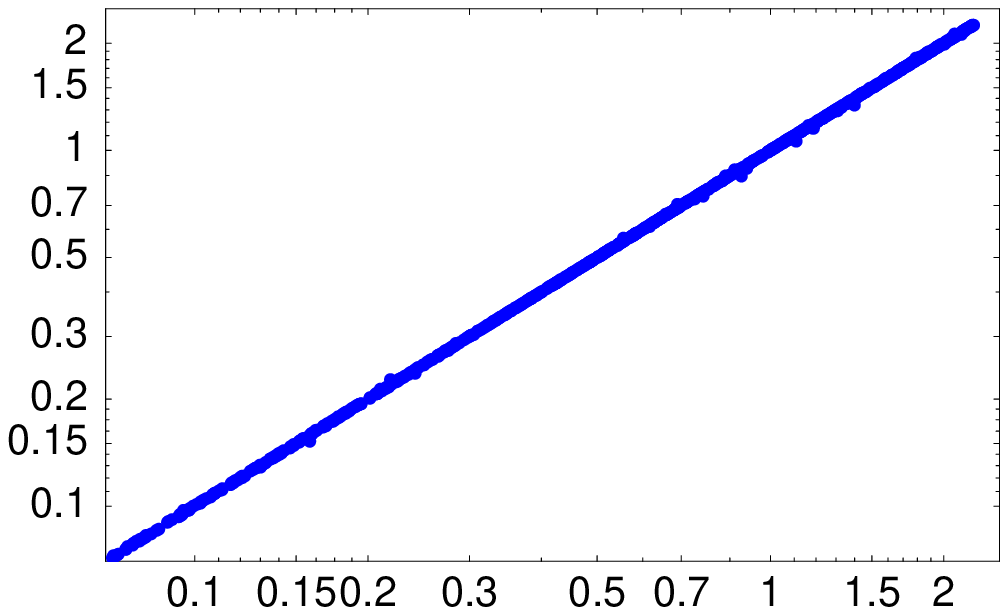}
\includegraphics[width=65mm,height=5cm]{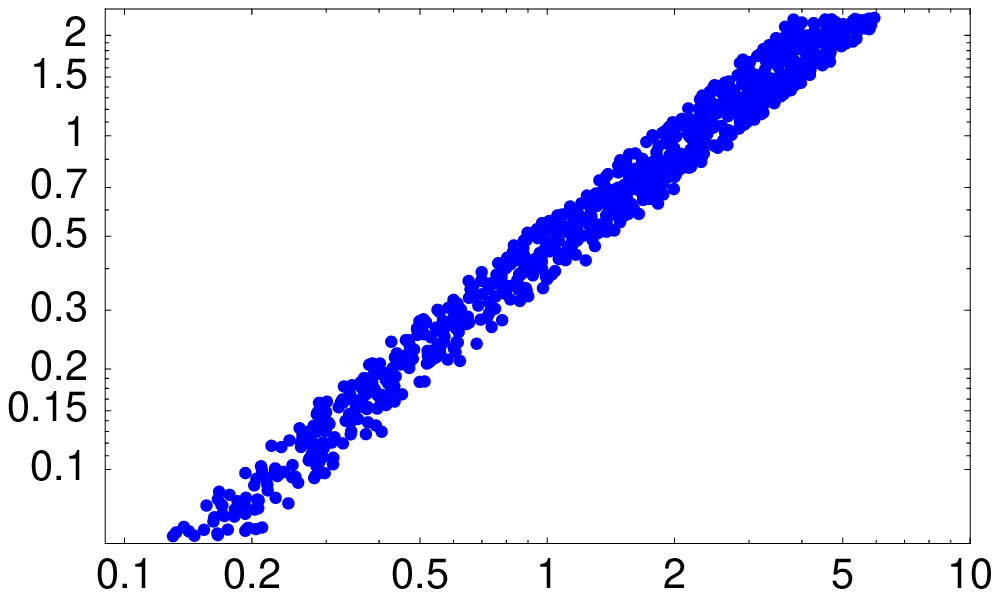}
\vskip-18mm\hskip5mm
\begin{rotate}{90}
$\frac{
Br({\tilde \nu}_e\to\tau\tau)/Br({\tilde \nu}_{\mu}\to\tau\tau)}
{Br({\tilde \nu}_e\to\mu\mu)/Br({\tilde \nu}_{\mu}\to e\mu)}$ 
\end{rotate}
\vskip12mm
\hskip55mm
$(\lambda_{133}/\lambda_{233})^2$ 
\hskip50mm $\tan^2\theta_{\odot}$
\vskip0mm

\caption{Double ratio of branching ratios $\frac{
Br({\tilde \nu}_e\to\tau\tau)/Br({\tilde \nu}_{\mu}\to\tau\tau)}
{Br({\tilde \nu}_e\to\mu\mu)/Br({\tilde \nu}_{\mu}\to e\mu)}$ 
as a function of a) (left) $(\lambda_{133}/\lambda_{233})^2$ 
and b) (right) $\tan^2\theta_{\odot}$ for scenario II. Top 
panel: All $\lambda_{ijk}$ same order of magnitude. Bottom panel: 
Assuming $\lambda_{133}$ and $\lambda_{233}$ larger than all other 
$\lambda_{ijk}$ by a factor of a few.}
\label{fig:solcaselam}
\end{figure}

In this scenario we assume that the atmospheric neutrino mass scale is 
due to tree-level (bilinear) R-parity violation, while the solar 
neutrino mass scale is due to charged scalar loops ($\lambda_{ijk}$), 
assuming $\lambda'_{ijk} \ll \lambda_{ijk}$. 
In this scenario the atmospheric neutrino data as well as $U^2_{e3}$
fix the allowed ranges for the $v_i$ whereas the solar neutrino data
fix the ranges for $\lambda_{133}$ and $\lambda_{233}$. 

We have calculated branching ratios for invisible sneutrino decays 
and find that they are smaller than for the bilinear-only 
case shown in Fig.~(\ref{fig:inv}), and thus not measurable. Also sneutrino 
decay lengths in this scenario are smaller than the ones shown 
in Fig.~(\ref{fig:len}). 
From a phenomenological point of view this scenario mainly differs 
from the bilinear-only case in the fact that (by assumption) there 
are very few jets in the decays of sneutrinos. 

Ratios of branching ratios can measure ratios of $\lambda_{ijk}$ and 
despite the fact that we assume here no hierarchy for the $\lambda_{ijk}$ 
this feature can be used to create a cross check of the scenario with 
solar neutrino physics. As an example we show in Fig.~(\ref{fig:solcaselam})
 double ratios of branching ratios as a 
function of $(\lambda_{133}/\lambda_{233})^2$ (left) and as a function 
of $\tan^2\theta_{\odot}$ (right). The correlation of this double 
ratio with $(\lambda_{133}/\lambda_{233})^2$ can be easily understood, 
since 

\begin{eqnarray}
\label{double}
Br({\tilde \nu}_e\to\tau\tau) \simeq c_{{\tilde \nu}_e} \lambda_{133}^2 
& \hskip5mm & Br({\tilde \nu}_{\mu}\to\tau\tau)\simeq c_{{\tilde \nu}_{\mu}} 
                                    \lambda_{233}^2 \\ \nn
Br({\tilde \nu}_e\to\mu\mu)  \simeq c_{{\tilde \nu}_e} \lambda_{122}^2 
& \hskip5mm & 
Br({\tilde \nu}_{\mu}\to e\mu) \simeq c_{{\tilde \nu}_{\mu}} 
                                    \lambda_{122}^2 
\end{eqnarray}
where $c_{{\tilde \nu}_{\mu}}$ and $c_{{\tilde \nu}_e}$ are constants 
which differ for the different generations of sneutrinos but 
which cancel in the double ratio. 
Here we have considered two limiting scenarios (i) $|\lambda_{133}|$ and 
$|\lambda_{233}|$ are larger than the other $\lambda_{ijk}$ and (ii) all 
of them are in the same order of magnitude. 
Interestingly, Fig. (\ref{fig:solcaselam}) demonstrates that a correlation 
of $(\lambda_{133}/\lambda_{233})^2$ with the solar angle exists even 
if all $\lambda_{ijk}$ are of the same order of magnitude. 
Clearly, these two scenarios can easily be distinguished at future 
collider experiments as in the first scenario $\tilde \nu_e$ and 
$\tilde \nu_\mu$ decay mainly into taus and $\tilde \nu_\tau$ decays 
mainly in a $l^\pm \tau^\pm$ (l=$e,\mu$) whereas in the second scenario 
the variety of different decay channels
is much larger for all sneutrinos.

\subsection{Scenario III: $v_{i} \simeq 0$}

In this scenario we assume that the neutrino mass matrix is 
dominated by trilinear terms and that the sneutrino 
vevs are negligible. There are thus no invisible decays of 
sneutrinos. We have checked that decay lengths in this scenario 
are even shorter than in scenario II.

\begin{figure}[t]
\hskip10mm
\includegraphics[width=65mm,height=5cm]{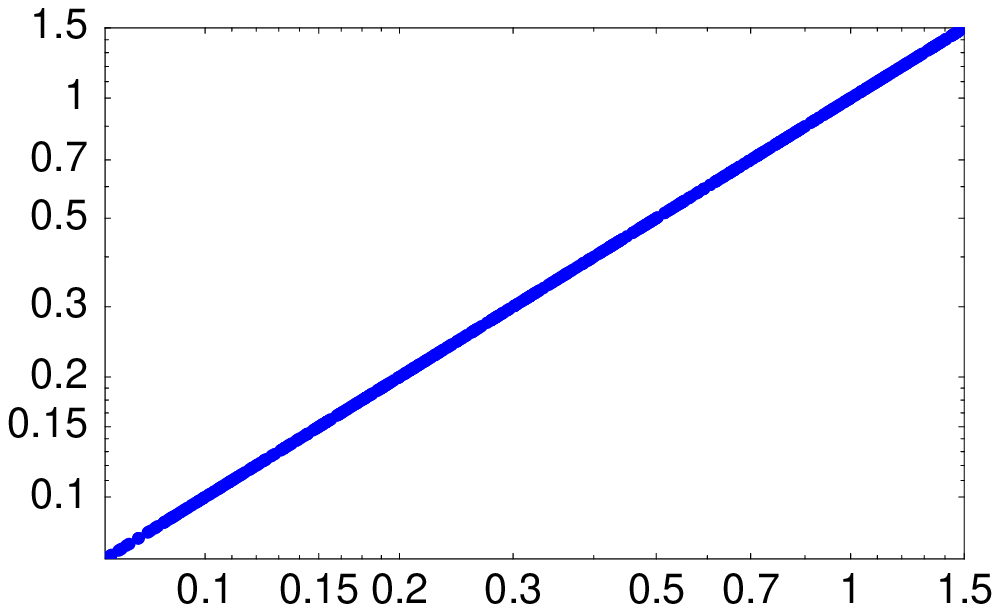}
\includegraphics[width=65mm,height=5cm]{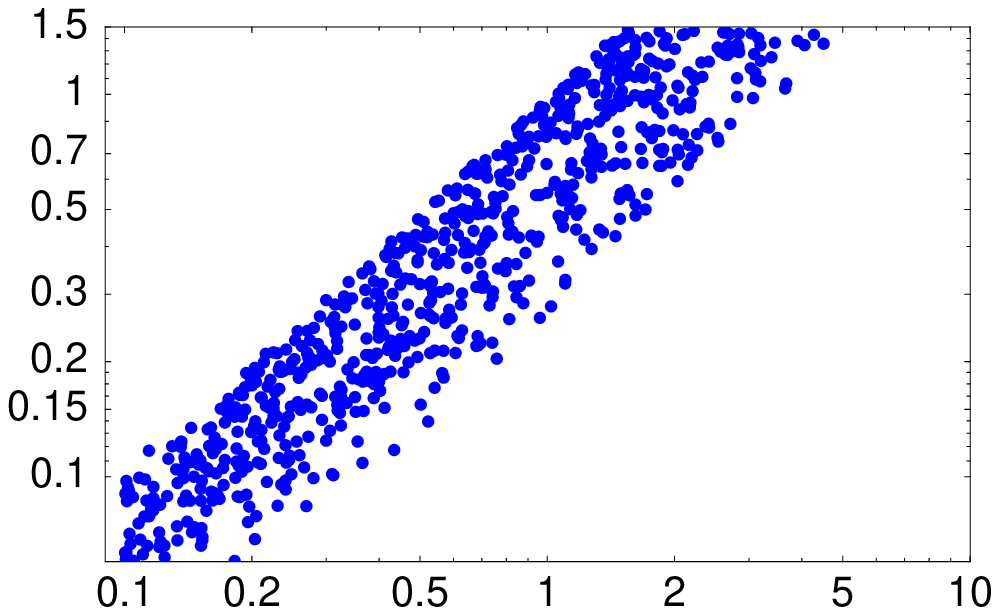}
\vskip-18mm\hskip5mm
\begin{rotate}{90}
$\frac{
Br({\tilde \nu}_e\to\tau\tau)/Br({\tilde \nu}_{\mu}\to\tau\tau)}
{Br({\tilde \nu}_e\to\mu\mu)/Br({\tilde \nu}_{\mu}\to e\mu)}$ 
\end{rotate}
\vskip12mm
\hskip55mm
$(\lambda_{133}/\lambda_{233})^2$ 
\hskip50mm $\tan^2\theta_{\odot}$
\vskip0mm

\hskip10mm
\includegraphics[width=65mm,height=5cm]{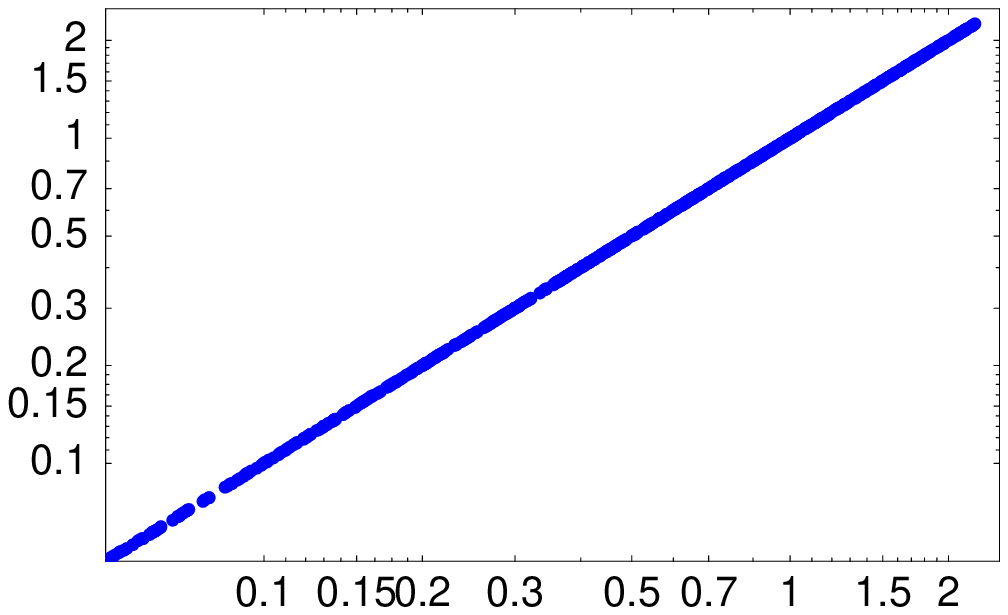}
\includegraphics[width=65mm,height=5cm]{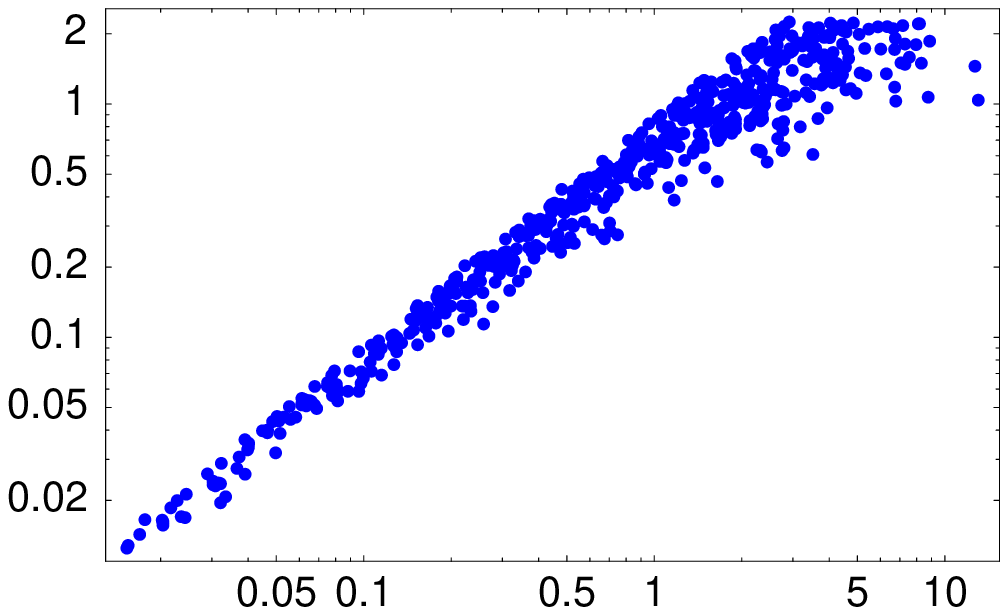}
\vskip-18mm\hskip5mm
\begin{rotate}{90}
$\frac{
Br({\tilde \nu}_{\mu}\to bb)/Br({\tilde \nu}_{\tau}\to b b)}
{Br({\tilde \nu}_{\mu}\to\tau\tau)/Br({\tilde \nu}_{\tau}\to \mu\tau)}$
\end{rotate}
\vskip12mm
\hskip55mm
$(\lambda'_{233}/\lambda'_{333})^2$
\hskip50mm $\tan^2\theta_{Atm}$
\vskip0mm

\caption{Top panel: Double ratio of branching ratios $\frac{
Br({\tilde \nu}_e\to\tau\tau)/Br({\tilde \nu}_{\mu}\to\tau\tau)}
{Br({\tilde \nu}_e\to\mu\mu)/Br({\tilde \nu}_{\mu}\to e\mu)}$ 
versus a) (left) $(\lambda_{133}/\lambda_{233})^2$ 
and b) (right) $\tan^2\theta_{\odot}$ for scenario III. 
Bottom panel: Double ratio of branching ratios $\frac{
Br({\tilde \nu}_{\mu}\to bb)/Br({\tilde \nu}_{\tau}\to b b)}
{Br({\tilde \nu}_{\mu}\to\tau\tau)/Br({\tilde \nu}_{\tau}\to \mu\tau)}$ 
versus a) (left) $(\lambda'_{233}/\lambda'_{333})^2$ 
and b) (right) $\tan^2\theta_{Atm}$. See text.} 
\label{fig:scenarioIV}
\end{figure}

Even for all $\lambda_{ijk}$ and $\lambda'_{ijk}$ different 
from zero ratios of $\lambda_{ijk}$ (and $\lambda'_{ijk}$) 
can be measured from double ratios of branching ratios as 
demonstrated in Fig. (\ref{fig:scenarioIV}), left panel.

If one assumes in addition (the validity of this assumption can be 
checked by measuring branching ratios) that $\lambda_{i33}$ and 
$\lambda'_{233}$ and $\lambda'_{333}$ are somewhat larger than 
the other $\lambda_{ijk}$ and $\lambda'_{ijk}$ one can make a 
consistency check with solar and atmospheric angle, as shown 
in Fig. (\ref{fig:scenarioIV}), right panel. Note that the 
correlation with the solar angle will be lost if all $\lambda_{ijk}$ 
are of the same size and/or $\lambda'_{ijk}$ with $j,k \ne 3$ 
are about equal to $\lambda_{i33}$ . Also the correlation with 
the atmospheric angle gets less pronounced if $\lambda'_{133}$ 
approaches $\lambda'_{333}$. However, the latter would be in 
contradiction to the observed smallness of the reactor angle and 
thus rule out one assumption of scenario III (atmospheric scale 
due to $\lambda'_{ijk}$) as well.

\section{Conclusion}
\label{sec:conc}

We have calculated the decay properties of sneutrinos assuming that 
they are the 
lightest supersymmetric particles in general R-parity violating 
models. The decay properties can be used to get information of the
relative importance of bilinear R-parity breaking couplings versus
trilinear R-parity breaking couplings which is quite interesting
from the model building point of view. We have seen that in the case
of bilinear dominance two important features occur:
(i) the existence of the invisible decay mode $\tilde \nu_i \to \nu_j \nu_k$
with a branching ratio of the order $10^{-4}$  
and (ii) the visible decay modes are governed by the lepton and down-quark
Yukawa couplings with the dominant mode into $\bar{b} b$ followed by 
$\tau^+ \tau^-$. In the case that trilinear terms become important both
features are changed as we have shown. 

Let us briefly comment on differences in the phenomenology
between neutralino
LSP, the most commonly studied LSP candidate, and 
sneutrino LSPs at future collider experiments such as LHC or a prospective ILC
in scenarios where R-parity is broken by lepton
number breaking terms. The most striking difference is that
sneutrino LSPs decay nearly to 100\% visible states with hardly any
missing energy whereas the neutralino has several final states containing
neutrinos. This implies that in the scenarios discussed here there are
several decay chains relevant for LHC which do not contain any missing
energy, e.g.~$\tilde q_L \to q' \tilde \chi^\pm_1 \to q' l^\pm \tilde \nu_l \to
 q' l^\pm b \bar{b}$ which clearly affects the strategies to identify the
particles as well as the determination of the underlying model parameters. 
However,
decay chains containing a neutralino still lead to the missing energy signature
in such cases due the the decay $\tilde \chi^0_i \to \nu \tilde \nu$, which,
depending on the mass difference $m_{\tilde \chi^0_i} - m_{\tilde \nu}$,
still can be sizable.

Moreover, we have shown that despite the large number of parameters 
certain ratios of branching ratios of sneutrino decays can be used 
to devise consistency checks with neutrino physics in all scenarios considered.
To accomplish this we have proposed to flavour tack the sneutrinos in chargino
decays as this substantially increases the possibilities to perform 
the cross check between observables in neutrino experiments and
observables at future high energy collider experiments.

\section*{Acknowledgments}
This work was supported by Spanish grant BFM2002-00345, by the
European Commission Human Potential Program RTN network
HPRN-CT-2000-00148 and by the European Science Foundation network
grant N.86.  M.H. and W.P. are supported by  MCyT Ramon y Cajal contracts.
W.P. was partly supported by the Swiss 'Nationalfonds'. D.A.S. is 
supported by a PhD fellowship by M.C.Y.T.

\end{document}